# Star-Trek Teleportation: A Possibility?

Kiang Wei Kho[1]

[1]School of Physics, National University of Ireland, Galway, Ireland

## Abstract

This paper describes a scheme, through which the quantum information as well as the structural information of a time-reversal invariant system can be teleported over a distance. I show that my teleportation scheme can be viewed as a form of reversible purification process by repeated interactions with an auxiliary quantum system.

e-mail: kiang.kho@nuigalway.ie/kho_97@yahoo.com

# I. INTRODUCTION

Quantum mechanics or QM has revolutionized our understanding of the universe since its discovery. An important consequence of QM is the Heisenberg's uncertainty principle (HUP), which states that no two non-commuting observable can be simultaneously measured with infinite accuracy. This flies in the face of classical Newtonian physics, which prescribes that every particle must have point-like behaviors. Fortunately, what was seemed to be an unbreakable barrier in the precision of all measuring apparatuses has surprisingly become the fundamental mechanism behind a new form of communication: quantum teleportation (or QT) [1]. As described in Bernett's landmark paper [2], the central element in QT is the concept of Einstein-Podolsky-Rosen or EPR entanglement, which was born out of an argument between Einstein and Bohr over the interpretation of HUP [3]. Such a peculiar quantum phenomenon has since been rigorously tested [4], eventually leading to such applications as encrypted transactions and high-fidelity quantum logic gates [5, 6]. Concomitantly, the complexity of the quantum system to be teleported has since also increased, from a simple photonic qubit system [2, 4] to a more complex many-body fermionic system, such as Bose-Einstein Condensate or BEC [7].

However, QT is not teleportation in the "Star-Trek" sense, since no disembodiment or reconstruction of any physical object actually occurs. What is actually being teleported in QT is the information about the quantum state of a physical system [2]. In the current paper, I propose a plausible scheme for transmitting both the q-information as well as the structural information of a quantum system over a distance. My theory can be generalized to accommodate molecular systems, in which case, the original system will undergo dissolution (*dematerialization*) at the sending end, and a replicate re-assembling (*rematerialization*) at the receiving end. As will be clear later, my teleportation scheme is essentially a form of reversible purification process, whereby the (quantum) system to be teleported (hereby referred to as the Alice's sample, or AS) is reversibly, and gradually, evolved into a specific, and manageable, eigenstate state through repeated interactions with a train of auxiliary particle pulses of lower dimension. During the purification process, the state of each of the auxiliary particles is first registered, after each interrogation, in a quantum memory at Alice's side, before being q-teleported or transmitted to Bob's memory. To reconstruct the original AS, Bob simply reverses the purification process using the recorded quantum states of the auxiliary particles, and performs on each of these states a specific symmetry transformation. Since, in this particular scheme, no *Bell* operator measurement is needed to be carried out on the AS, my teleportation scheme is much simplified and is thus, in principle, amenable to any AS, regardless of its complexity as well as phase (i.e. molecule, gas, liquid etc). Additionally, my technique generally does not impose a limitation on the maximum allowable thermal energies of the AS, i.e. the AS needs not be very "cold". This is in great contrast to the previous scheme of photon-mediated teleportation of matter-waves, where an efficient teleportation occurs only within a restricted range of mean particle velocity [8]. The current proposed scheme therefore put Star-Trek Teleportation (STT) much closer to the reality. Finally, I will comment on the relevancy of some of the currently existing quantum technologies to STT and suggest a plausible implementation setup based on the current model. This paper is divided into two essential parts: Section II and III discuss the general model of concatenated teleportation via purification process, and its classification, in the context of STT; in Section IV, I apply my theory in the STT of a physical object.

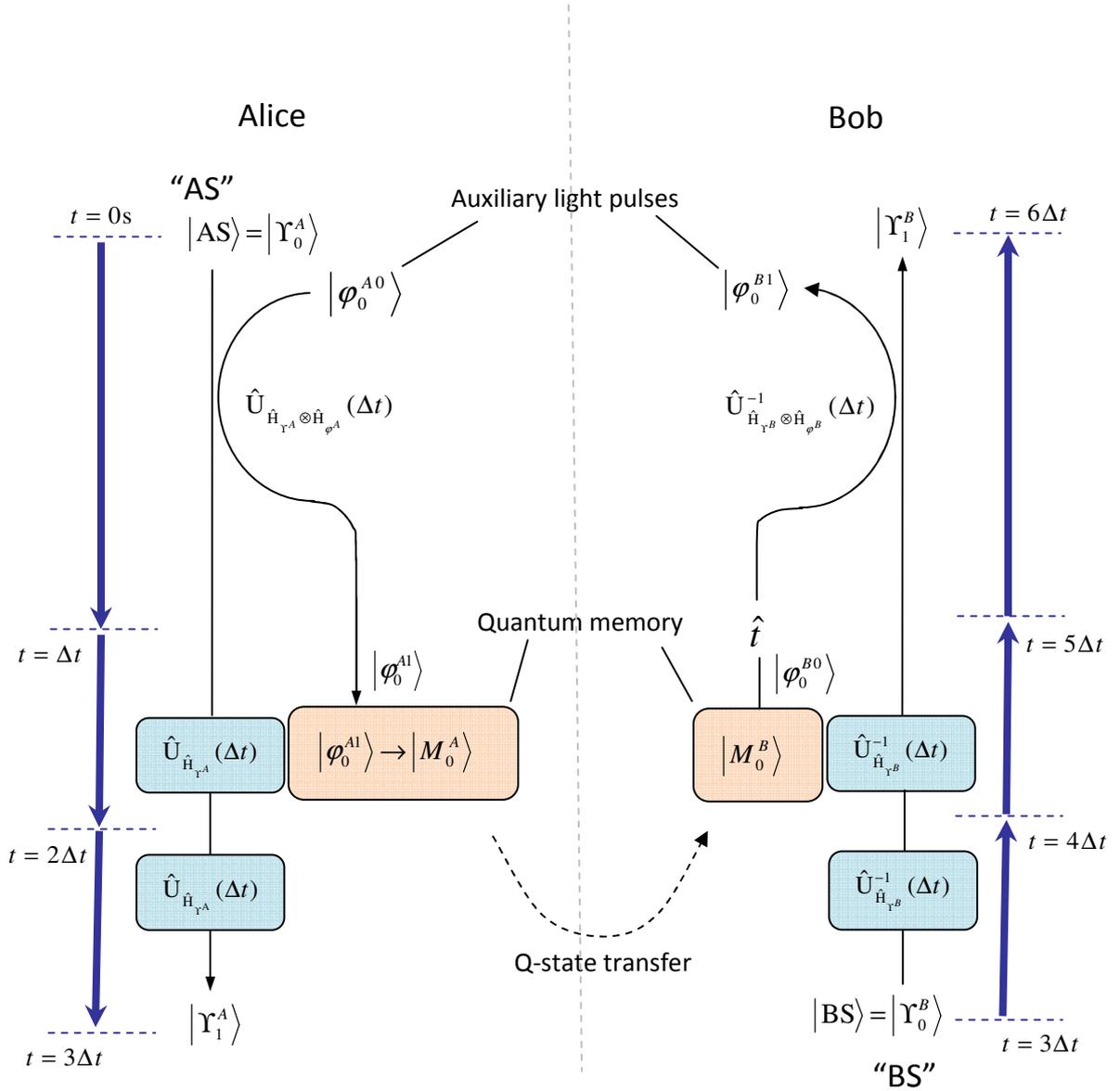

FIG. 1. A general teleportation scheme

## II. CONCATENATED-TELEPORTATION OF A COMPLEX QUANTUM SYSTEM

I shall begin with a general teleportation scheme as depicted in Fig. 1, whereby Alice intends to teleport or map an unknown state $|AS\rangle = |\Upsilon_0^A\rangle$ (spatial state, momentum state, etc) of the AS onto the state of the object that Bob has at hand (referred to as Bob's sample or BS). Let $|\Upsilon_0^A\rangle \in H^{N_{\Upsilon^A}}$ and $|\Upsilon_0^B\rangle \in H^{N_{\Upsilon^B}}$, where $H^{N_{\Upsilon^A}}$ and $H^{N_{\Upsilon^B}}$ are Hilbert spaces with dimensions $N_{\Upsilon^A}$ and $N_{\Upsilon^B}$, respectively. Let $|e_i^{\Upsilon^A}\rangle$, $i = 1,...,N_{\Upsilon^A}$ and $|e_j^{\Upsilon^B}\rangle$, $j = 1,...,N_{\Upsilon^B}$, be the orthogonal basis in $H^{N_{\Upsilon^A}}$ and $H^{N_{\Upsilon^B}}$, respectively. Let's also assume that both the AS and the BS are complex N-body fermionic systems consisting of the same atomic specie, and that $N_{\Upsilon^A} = N_{\Upsilon^B}$. I will also assume that the entire teleportation process is to be carried out in a piece-wise manner, i.e. in steps of time-interval $\Delta t$ (see lower panel of Fig. 1).

Supposed Alice begins the teleportation process by first interrogating the state $|AS\rangle$ with a single auxiliary light pulse of state $|\varphi_0^{A0}\rangle$ ($|\varphi_0^{A0}\rangle \in H^{N_{\varphi^A}}$) under the unitary operation $\hat{U}^{-1}_{\hat{H}_\Upsilon \otimes \hat{H}_\varphi}(\Delta t)$, in which $\hat{H}_\Upsilon \otimes \hat{H}_\varphi$ is the AS-auxiliary Hamiltonian. Note that I shall assume the auxiliary pulse to be of a lower dimension, i.e. $N_{\varphi^A} \ll N_{\Upsilon^A}$, and has a pulse-width of $\Delta t$. In the subsequent $\Delta t$ interval following the first interrogation, Alice stores the final state ($|\varphi_0^{A1}\rangle$) of the light pulse in her quantum memory $|M_0^A\rangle$ (i.e. $|\varphi_0^{A1}\rangle \to |M_0^A\rangle$), which has a coherence time significantly longer than the time needed for the entire teleportation process. During this $\Delta t$ interval, the state $|AS\rangle$ is allowed to evolve naturally under a unitary transformation $\hat{U}_{\hat{H}_\Upsilon}(\Delta t) = e^{-i\hat{H}_\Upsilon(\Delta t)}$, where $\hat{H}_\Upsilon$ is the Hamiltonian of the AS. In the interval $[2\Delta t, 3\Delta t]$ (see Fig. 1), the state of Alice's memory is q-teleported to (or mapped onto) Bob's memory; $|M_0^A\rangle \to |M_0^B\rangle$, during which time, $|AS\rangle$ is again allowed to evolved, eventually to a final state $|\Upsilon_1^A\rangle$. Supposed now that Bob wishes to re-create the original state $|\Upsilon_0^A\rangle$, by reversing all the operations carried out by Alice. Bob shall regard the reconstruction process a success, only if the coefficients for the final state $|BS\rangle = |\Upsilon_1^B\rangle$ are the same as those for the original state $|AS\rangle = |\Upsilon_0^A\rangle$, when expressed as linear superposition of their own bases. To begin the reconstruction process, Bob prepares BS with some initial state $|BS\rangle = |\Upsilon_0^B\rangle$, and acts on it an inverse unitary transformation $\hat{U}^{-1}_{\hat{H}_\Upsilon}(\Delta t)$, where $\hat{H}_\Upsilon$ is the Hamiltonian of the

BS. In the interval $[4\Delta t, 5\Delta t]$, Bob maps the state $|M_0^B\rangle$ of his quantum memory onto the state $|\varphi_0^{B0}\rangle$ (which spans the Hilbert space $H^{N_{\varphi^B}}$) of his own auxiliary light pulse, and performs on this particular light pulse a symmetry transformation $\hat{t}$. As in the case for $|\Upsilon_0^A\rangle$ and $|\Upsilon_1^B\rangle$, $|\varphi_0^{B0}\rangle$ and $|\varphi_0^{A1}\rangle$ are considered equal if their corresponding coefficients are the same with respect to their own bases. Bob again let $|BS\rangle$ evolves naturally during this time interval under the inverse unitary transformation $\hat{U}_{\hat{H}_\Upsilon}^{-1}(\Delta t)$. In the final interval $[5\Delta t, 6\Delta t]$, both the light pulse and the evolved state $\hat{U}_{\hat{H}_\Upsilon}^{-1}(2\Delta t)|\Upsilon_0^B\rangle$ interacts under the action of the unitary $\hat{U}_{\hat{H}_\Upsilon \otimes \hat{H}_\varphi}^{-1}(\Delta t)$. There is no doubt that if Bob could prepare the initial state $|BS\rangle = |\Upsilon_0^B\rangle$ wisely, $|\Upsilon_0^A\rangle$ shall be recreated in the final state $|\Upsilon_1^B\rangle$ of the BS. The question here is therefore not so much as to whether Bob can achieve the re-creation, but as to what is the set of "teleportable" $|\Upsilon_0^A\rangle$s - and ultimately how to expand this set to accommodate more-general states - if there exists only a classical channel for Alice to convey to Bob information about the final state $|\Upsilon_1^A\rangle$ of the AS, owing, perhaps, to the difficulties in obtaining a suitable EPR-pair consisting of members with high enough dimension (at least $N_{\Upsilon^A}$), or in carrying out a *Bell*-state measurement on the AS-auxiliary system.

To see this, I shall begin by stating the following bijective mapping for the eigenstates, $\hat{Q}_\Upsilon : |e_i^{\Upsilon^A}\rangle \mapsto |e_i^{\Upsilon^B}\rangle$, $\hat{Q}_\varphi : |e_j^{\varphi^A}\rangle \mapsto |e_j^{\varphi^B}\rangle$, $\forall i = 1,...,N_\Upsilon$, $\forall j = 1,...,N_\varphi$, which implies equivalency between $|e_i^{\Upsilon^A}\rangle$ (resp. $|e_j^{\varphi^A}\rangle$), and $|e_i^{\Upsilon^B}\rangle$ (resp. $|e_j^{\varphi^B}\rangle$), in which it is assumed that $N_{\Upsilon^A} = N_{\Upsilon^B} = N_\Upsilon$, $N_{\varphi^A} = N_{\varphi^B} = N_\varphi$, and $N_\varphi \ll N_\Upsilon$. With this proposition, it thus follows that $|\Upsilon^A\rangle = \sum_{i=1}^{N_\Upsilon} a_i |e_i^{\Upsilon^A}\rangle$ (resp. $|\varphi^A\rangle = \sum_{j=1}^{N_\varphi} b_j |e_j^{\varphi^A}\rangle$) and

$\left| \Upsilon^B \right\rangle = \sum_{i=1}^{N_\Upsilon} a_i \left| e_i^{\Upsilon^B} \right\rangle$ (resp. $\left| \varphi^B \right\rangle = \sum_{j=1}^{N_\varphi} b_j \left| e_j^{\varphi^B} \right\rangle$) will be considered equivalent. Now, I take

$\left| \Upsilon_0^A \right\rangle = \sum_{i=1}^{N_\Upsilon} a_i \left| e_i^{\Upsilon^A} \right\rangle$, and $\left| \varphi_0^{A0} \right\rangle = \sum_{j=1}^{N_\varphi} b_j \left| e_j^{\varphi^A} \right\rangle$, and express the combined state $\left| \Upsilon_0^A \otimes \varphi_0^{A0} \right\rangle$

(which is separable) as,

$$\begin{aligned} \left| \Upsilon_0^A \otimes \varphi_0^{A0} \right\rangle &= \left( \sum_{m=1}^{N_\Upsilon} a_m \left| e_m^{\Upsilon^A} \right\rangle \right) \left( \sum_{n=1}^{N_\varphi} b_n \left| e_n^{\varphi^A} \right\rangle \right) \\ &= \sum_{m,n=1}^{N_\Upsilon, N_\varphi} a_m b_n \left| e_m^{\Upsilon^A} \right\rangle \left| e_n^{\varphi^A} \right\rangle \\ &= \sum_{i,j=1}^{N_\Upsilon, N_\varphi} \sum_{m,n=1}^{N_\Upsilon, N_\varphi} \left[ a_m b_n \left\langle \Phi_{ij}^A \middle| e_m^{\Upsilon^A} \right\rangle \left| e_n^{\varphi^A} \right\rangle \right] \left| \Phi_{ij}^A \right\rangle \end{aligned} \qquad (1)$$

where $a_m, b_n \in \mathbb{C}$, $\sum_{m=1}^{N_\Upsilon} |a_m|^2 = 1$, $\sum_{n=1}^{N_\varphi} |b_n|^2 = 1$.

From Eqn. 1 and according to Fig. 1, I obtain for the state $\left| \Upsilon_1^A \otimes M_0^B \right\rangle$ at $t = 3\Delta t$ as,

$$\begin{aligned} \left| \Upsilon_1^A \otimes M_0^B \right\rangle &= \left( \hat{U}_{\hat{H}_{\Upsilon^A}}(2\Delta t) \otimes \hat{I} \right) \hat{T}_{A \to B} \hat{M}_{\varphi \to M} \hat{U}_{\hat{H}_{\Upsilon^A} \otimes \hat{H}_{\varphi^A}}(\Delta t) \left| \Upsilon_0^A \otimes \varphi_0^{A0} \right\rangle \\ &= \sum_{i,j=1}^{N_\Upsilon, N_\varphi} \left( \sum_{m,n=1}^{N_\Upsilon, N_\varphi} a_m b_n \left\langle \Phi_{ij}^A \middle| e_m^{\Upsilon^A} \right\rangle \left| e_n^{\varphi^A} \right\rangle \right) \left| e_i^{\Upsilon^A} \right\rangle \left| e_j^{M^B} \right\rangle \end{aligned} \qquad (2)$$

in which, I have defined

$$\left| \Phi_{ij}^A \right\rangle = \left[ \left( \hat{U}_{\hat{H}_{\Upsilon^A}}(2\Delta t) \otimes \hat{I} \right) \hat{T}_{A \to B} \hat{M}_{\varphi \to M} \hat{U}_{\hat{H}_{\Upsilon^A} \otimes \hat{H}_{\varphi^A}}(\Delta t) \right]^{-1} \left| e_i^{\Upsilon^A} \right\rangle \left| e_j^{M^B} \right\rangle,$$

where $\hat{U}_{\hat{H}_{\Upsilon^A} \otimes \hat{H}_{\varphi^A}}(\Delta t) = e^{-i\left[ \hat{H}_{\Upsilon^A} \otimes \hat{H}_{\varphi^A} \right](\Delta t)}$, and $\hat{H}_{\Upsilon^A} \otimes \hat{H}_{\varphi^A}$ is the total Hamiltonian for the combined system $\left| \Upsilon^A \otimes \varphi^{A0} \right\rangle$, and $\hat{M}_{\varphi \to M}$ represents a bijective mapping of the auxiliary space $\left\{ \left| \varphi_0^{A1} \right\rangle \right\}$ onto the memory space $\left\{ \left| M_0^A \right\rangle \right\}$; $\hat{M}_{\varphi \to M} : \left\{ \left| \Upsilon \otimes \varphi \right\rangle \right\} \to \left\{ \left| \Upsilon \otimes M \right\rangle \right\}$, and

$\hat{T}_{A \to B}$ Alice's memory space $\{|M_0^A\rangle\}$ onto Bob's memory space $\{|M_0^B\rangle\}$; to

$$\hat{T}_{A \to B} : \{|\Upsilon \otimes M^A\rangle\} \to \{|\Upsilon \otimes M^B\rangle\}.$$

From Eqn. 2, I can write $|\Upsilon_1^A\rangle$ as,

$$|\Upsilon_1^A\rangle = \sum_{i=1}^{N_\Upsilon} \left( \sum_{j=1}^{N_\varphi} \sum_{m,n=1}^{N_\Upsilon, N_\varphi} a_m b_n \langle \Phi_{ij}^A | e_m^{\Upsilon^A} \rangle | e_n^{\varphi^A} \rangle \right) | e_i^{\Upsilon^A} \rangle \tag{3}$$

In order for Bob to reconstruct the original state $|\Upsilon_0^A\rangle$ under an inverse unitary operation, he must bijectively map $|\Upsilon_1^A\rangle$ onto $|\Upsilon_0^B\rangle$; $|\Upsilon_1^A\rangle \mapsto |\Upsilon_0^B\rangle$, and prepares $|\Upsilon_0^B\rangle$ as,

$$|\Upsilon_0^B\rangle = \sum_{i=1}^{N_\Upsilon} \left( \sum_{j=1}^{N_\varphi} \sum_{m,n=1}^{N_\Upsilon, N_\varphi} a_m b_n \langle \Phi_{ij}^A | e_m^{\Upsilon^A} \rangle \otimes | e_n^{\varphi^A} \rangle \right) | e_i^{\Upsilon^B} \rangle, \tag{4}$$

or more precisely, he must prepare $|\Upsilon_0^B \otimes M_0^B\rangle$ as the following,

$$|\Upsilon_0^B \otimes M_0^B\rangle = \sum_{i,j=1}^{N_\Upsilon, N_\varphi} \left( \sum_{m,n=1}^{N_\Upsilon, N_\varphi} a_m b_n \langle \Phi_{ij}^A | e_m^{\Upsilon^A} \rangle | e_n^{\varphi^A} \rangle \right) | e_i^{\Upsilon^B} \rangle | e_j^{M^B} \rangle \tag{5}$$

While the satisfaction of Eqn. 5 will always lead to a successful reconstruction of $|\Upsilon_0^A\rangle$ in BS, Eqn. 5 is, however, fairly general, and does not take into consideration of the fact that both Alice and Bob are spatially well separated. In other word, Eqn. 5 may demand that the states $|\Upsilon_0^A\rangle$, $|\Upsilon_0^B\rangle$ and $|\Upsilon_1^A\rangle$ be entangled – an impractical requirement for Alice and Bob. To see this, I express, by virtue of Eqn. 1 and 4, the relationship between $|\Upsilon_0^A\rangle$ and $|\Upsilon_0^B\rangle$ as,

$$|\Upsilon_0^A \otimes \Upsilon_0^B\rangle = \sum_{m,i=1}^{N_\Upsilon} a_m \left( \sum_{n,j=1}^{N_\varphi} b_n \langle \Phi_{ij}^A | e_m^{\Upsilon^A} \rangle | e_n^{\varphi^A} \rangle \right) | e_m^{\Upsilon^A} \rangle | e_i^{\Upsilon^B} \rangle \tag{6}$$

through which, one can easily see that $|\Upsilon_0^A\rangle$ and $|\Upsilon_0^B\rangle$ are generally entangled. If now the separablity between $|\Upsilon_0^A\rangle$ and $|\Upsilon_0^B\rangle$ is assumed, i.e.

$$a_m \left( \sum_{n,j=1}^{N_\varphi} b_n \langle \Phi_{ij}^A | e_m^{\Upsilon^A} \rangle | e_n^{\varphi^A} \rangle \right) = f_m g_i \qquad (7)$$

and re-writing Eqn. 1 as,

$$|\Upsilon_0^A \otimes \varphi_0^A\rangle = \sum_{i,j=1}^{N_\Upsilon, N_\varphi} \alpha_{ij} |\Phi_{ij}^A\rangle$$

, where $\sum_{m,n=1}^{N_\Upsilon, N_\varphi} a_m b_n \langle \Phi_{ij}^A | e_m^{\Upsilon^A} \rangle \otimes | e_n^{\varphi^A} \rangle = \alpha_{ij}$, I can obtain for $|\Upsilon_1^A \otimes \Upsilon_0^B\rangle$,

$$|\Upsilon_1^A \otimes \Upsilon_0^B\rangle = \sum_{i,j=1}^{N_\Upsilon} \left( \sum_{l=1}^{N_\varphi} \alpha_{il} \alpha_{jl} \right) |e_i^{\Upsilon^A}\rangle |e_j^{\Upsilon^B}\rangle \qquad (8)$$

By comparing Eqn. 8 with Eqn. 6, one can further see that although,

$$\sum_{j=1}^{N_\varphi} \sum_{m,n=1}^{N_\Upsilon, N_\varphi} \{ a_m b_n \langle \Phi_{ij}^A | \} | e_m^{\Upsilon^A} \rangle | e_n^{\varphi^A} \rangle = \left( \sum_{m=1}^{N_\Upsilon} f_m \right) g_i = \sum_{j=1}^{N_\varphi} \alpha_{ij}$$

, there is no implication that $\alpha_{ij} = \alpha_i \beta_j$ is generally valid; therefore, $|\Upsilon_1^A\rangle$ and $|\Upsilon_0^B\rangle$ are generally entangled, even when $|\Upsilon_0^A\rangle$ and $|\Upsilon_0^B\rangle$ are separable.

To this end, and given the fact that Alice and Bob are separated, and that $|\Upsilon_1^A\rangle$ and $|\Upsilon_0^B\rangle$ are too complex a state to be made entangled, I shall thereafter consider only the particular case, where $|\Upsilon_1^A\rangle$ and $|\Upsilon_0^B\rangle$ are separable (i.e. $\alpha_{ij} = \alpha_i \beta_j$). This concomitantly suggests that $|\Upsilon_1^A \otimes M_0^B\rangle$, and hence $|\Upsilon_0^B \otimes M_0^B\rangle$, be separable as well. Hence, one can write, for this particular case, for $|\Upsilon_0^B \otimes M_0^B\rangle$ as,

$$|\Upsilon_0^B \otimes M_0^B\rangle = \sum_{i,j=1}^{N_\Upsilon, N_\varphi} \alpha_i \beta_j |e_i^{\Upsilon^B}\rangle |e_j^{M^B}\rangle$$

Accordingly, the reconstruction process carried out by Bob (see Fig. 1) is therefore,

$$\left|\Upsilon_1^B \otimes \varphi_0^{B1}\right\rangle = \hat{U}_{\hat{H}_{\Upsilon^B} \otimes \hat{H}_{\varphi^B}} (\Delta t)^{-1} \hat{t}_\varphi \hat{M}_{\varphi \to M}^{-1} \left(\hat{U}_{\hat{H}_{\Upsilon^B}}(2\Delta t) \otimes \hat{I}\right)^{-1} \left|\Upsilon_0^B \otimes M_0^B\right\rangle$$

$$= \sum_{i,j=1}^{N_\Upsilon, N_\varphi} \alpha_i \beta_j \left|\Phi_{ij}^B\right\rangle \quad (9)$$

$$= \sum_{m,n=1}^{N_\Upsilon, N_\varphi} \sum_{i,j=1}^{N_\Upsilon, N_\varphi} \left[\alpha_i \beta_j \left\langle e_m^{\Upsilon^B}\right|\left\langle e_n^{\varphi^B}\Big|\Phi_{ij}^B\right\rangle\right]\left|e_m^{\Upsilon^B}\right\rangle\left|e_n^{\varphi^B}\right\rangle$$

For simplicity, I shall set $\hat{t}_\varphi = \hat{I}$ for the moment. By comparing

$$\left|\Upsilon_0^A \otimes \varphi_0^{A0}\right\rangle = \sum_{i,j=1}^{N_\Upsilon, N_\varphi} \alpha_{ij} \left|\Phi_{ij}^A\right\rangle = \sum_{i,j=1}^{N_\Upsilon, N_\varphi} \alpha_i \beta_j \left|\Phi_{ij}^A\right\rangle, \text{ with Eqn. 1, I obtain,}$$

$$a_m b_n = \sum_{i,j=1}^{N_\Upsilon, N_\varphi} \left\{\alpha_i \beta_j \left\langle e_m^{\Upsilon^A}\right|\left\langle e_n^{\varphi^A}\right|\right\}\left|\Phi_{ij}^A\right\rangle \quad (10)$$

Thus, from Eqn. 9, I can write,

$$\left|\Upsilon_1^B\right\rangle = \sum_{m,n=1}^{N_\Upsilon, N_\varphi} \sum_{i,j=1}^{N_\Upsilon, N_\varphi} \left\{\alpha_i \beta_j \left\langle e_m^{\Upsilon^B}\right|\left\langle e_n^{\varphi^B}\Big|\Phi_{ij}^B\right\rangle\right\}\left|e_m^{\Upsilon^B}\right\rangle$$

$$= \sum_{m=1}^{N_\Upsilon} a_m \left(\sum_{n=1}^{N_\varphi} b_n\right)\left|e_m^{\Upsilon^B}\right\rangle \quad (11)$$

Based on the above definition on the equality between the eigenstates $\left|e_i^{\Upsilon^A}\right\rangle$ and $\left|e_i^{\Upsilon^B}\right\rangle$, Eqn. 11 therefore shows that $\left|\Upsilon_1^B\right\rangle$ can still be a perfect reconstruction of $\left|\Upsilon_0^A\right\rangle$, even when both AS and BS (i.e. $\left|\Upsilon_1^A\right\rangle$ and $\left|\Upsilon_0^B\right\rangle$) are never entangled.

While Eqn. 11 does not appear much as a surprise, its derivation has implicitly assumed that Bob can precisely "duplicate" the state $\left|\Upsilon_1^A\right\rangle$ in the BS prior to the reconstruction process, (see Eqn. 4). This assumption is however too strong, and certainly goes against the no-cloning theorem. In addition, as has assumed earlier on, there is also no quantum channel available to permit Alice to precisely map the state $\left|\Upsilon_1^A\right\rangle$ onto $\left|BS\right\rangle$. Alice, however, is allowed to communicate with Bob about the state $\left|\Upsilon_1^A\right\rangle$ via a classical channel. To achieve this, Alice will have to perform a classical measurement ($\hat{P}_p$) on $\left|\Upsilon_1^A\right\rangle$, and convey to Bob the outcome, so that Bob can begin the reconstruction process with his BS prepared with an appropriate eigenstate. Assuming, the final projected AS state be $\left|e_p^{\Upsilon^A}\right\rangle$, i.e.

$$\hat{P}_p : \left|\Upsilon_1^A\right\rangle = \sum_{i=1}^{N_\Upsilon} k_i \left|e_i^{\Upsilon^A}\right\rangle \rightarrow \left|e_p^{\Upsilon^A}\right\rangle, \quad \text{(where } k_p \neq 0\text{)}$$

and Bob prepares the initial $\left|BS\right\rangle$ to be $\left|e_p^{\Upsilon^B}\right\rangle$. Eqn. 11 then becomes,

$$\begin{aligned}\left|\Upsilon_1^B\right\rangle &= \sum_{m,n=1}^{N_\Upsilon,N_\varphi} \sum_{i,j=1}^{N_\Upsilon,N_\varphi} \left\{\delta_{ip}\beta_j \left\langle e_m^{\Upsilon^B}\right|\left\langle e_n^{\varphi^B}\right|\Phi_{ij}^B\right\rangle\right\}\left|e_m^{\Upsilon^B}\right\rangle \\ &= \sum_{m,n=1}^{N_\Upsilon,N_\varphi} \sum_{j=1}^{N_\varphi} \left\{\beta_j \left\langle e_m^{\Upsilon^B}\right|\left\langle e_n^{\varphi^B}\right|\Phi_{pj}^B\right\rangle\right\}\left|e_m^{\Upsilon^B}\right\rangle\end{aligned} \quad (12)$$

The fidelity of the AS's original state ($\left|AS\right\rangle = \left|\Upsilon_0^A\right\rangle$) being reconstructed in $\left|\Upsilon_1^B\right\rangle$ can be quantified as,

$$\left\langle\Upsilon_0^A\middle|\Upsilon_1^B\right\rangle = \sum_{m=1}^{N_\Upsilon} a_m^* \sum_{n,j=1}^{N_\varphi} \left\{\beta_j \left\langle e_m^{\Upsilon^B}\right|\left\langle e_n^{\varphi^B}\right|\right\}\Phi_{pj}^B\right\rangle \quad (13)$$

From Eqn. 13, it can be seen that the teleportation is successful, i.e. $\left\langle\Upsilon_A^0\middle|\Upsilon_B^1\right\rangle$ equal to unity, only if,

$$\sum_{n,j=1}^{N_\varphi} \beta_j \left\langle e_m^{\Upsilon^B}\right|\left\langle e_n^{\varphi^B}\right|\Phi_{pj}^B\right\rangle = a_m \quad (14)$$

Since there could, at most, be only $N_\varphi$ non-trivial $b_n$s, the set of "teleportable" state $\left|\Upsilon_0^A\right\rangle^t$ that satisfies Eqn. 14, for any given state $\left|\varphi_0^{A0}\right\rangle$, and any given projection measurement $\hat{P}_p$ constitutes only a $N_\varphi$-dimensional sub-space $\left\{\left|\Upsilon_0^A\right\rangle^t\right\}\bigg|_{\hat{P}_p,\left|\varphi_0^{A0}\right\rangle} \subset H^{N_\Upsilon}$. This stems simply

from the fact that the limited dimension of the auxiliary space ($|\varphi_0^A\rangle$ and $|\varphi_0^B\rangle$) prohibits the realization of a general teleportation.

If however Alice can repeatedly probe $|AS\rangle$ for at least $N'$ times with a sequence of auxiliary light pulses, such that $N'N_\varphi \geq N_\Upsilon$, Alice could still send (a general) $|\Upsilon_0^A\rangle$ to Bob. In this particular case, each AS-light-pulse interaction will partially evolve $|AS\rangle$ toward a targeted eigenstate. This is known as concatenated teleportation [9], and is depicted in Fig. 2 as a cascade series of partial teleportation steps. The entire teleportation process can be described as,

$$\left|\Upsilon_{N'}^B \otimes \prod_{i=0}^{N'-1}\varphi_i^{B1}\right\rangle = \left(\hat{O}_B \hat{M}_{B,M\to\varphi}\right)\hat{P}_p \hat{T}_{A\to B}\left(\hat{M}_{A,\varphi\to M}\hat{O}_A\right)\left|\Upsilon_0^A \otimes \prod_{i=0}^{N'-1}\varphi_i^{A0}\right\rangle \quad (15)$$

in which, $\left|\Upsilon_0^A \otimes \varphi_0^{A0} \otimes \varphi_1^{A0} \otimes ... \otimes \varphi_{N'-2}^{A0} \otimes \varphi_{N'-1}^{A0}\right\rangle = \left|\Upsilon_0^A \otimes \prod_{i=0}^{N'-1}\varphi_i^{A0}\right\rangle$, $\hat{M}_{A,\varphi\to M}$ (resp. $\hat{M}_{B,M\to\varphi}$) bijectively maps $\left|\prod_{i=0}^{N'-1}\varphi_i^{A0}\right\rangle$ onto $\left|\prod_{i=0}^{N'-1}M_i^{A0}\right\rangle$ (resp. $\left|\prod_{i=0}^{N'-1}M_i^{B0}\right\rangle$ onto $\left|\prod_{i=0}^{N'-1}\varphi_i^{B0}\right\rangle$), and ,

$$\hat{O}_A = \left(_{N'}\hat{R}\left(\hat{U}_{\hat{H}_{\Upsilon^A}}(2\Delta t)\otimes\hat{I}\right)\hat{U}_{\hat{H}_{\Upsilon^A}\otimes\hat{H}_{\varphi^A}}(\Delta t)\right)^{N'}, \quad \hat{O}_B = \left(\hat{U}_{\hat{H}_{\Upsilon^B}\otimes\hat{H}_{\varphi^B}}^{-1}(\Delta t)\left(\hat{U}_{\hat{H}_{\Upsilon^B}}^{-1}(2\Delta t)\otimes\hat{I}\right)_{N'}\hat{R}^{-1}\right)^{N'}$$ and

$_{N'}\hat{R}$ defines an operator that left-rotates $\varphi_0^{A0} \otimes \varphi_1^{A0} \otimes ... \otimes \varphi_{N'-2}^{A0} \otimes \varphi_{N'-1}^{A0}$;

$$_{N'}\hat{R}\left|\Upsilon_0^A \otimes \varphi_0^{A0} \otimes \varphi_1^{A0} \otimes ... \otimes \varphi_{N'-1}^{A0}\right\rangle = \left|\Upsilon_0^A \otimes \varphi_1^{A0} \otimes \varphi_2^{A0} \otimes ... \otimes \varphi_{N'-1}^{A0} \otimes \varphi_0^{A0}\right\rangle$$

, and the dimension of $\left|\prod_{i=0}^{N'-1}\varphi_i^{A0}\right\rangle$ is $N'N_\varphi$.

As has discussed above, and implied by Eqn. 13 and 14, the pre-requisite for a successful teleportation is that Alice be able to project her AS onto an eigenstate $\left|e_p^{\Upsilon^A}\right\rangle$, i.e.

$$\hat{O}_A\left|\Upsilon_0^A \otimes \prod_{i=0}^{N'-1}\varphi_i^{A0}\right\rangle = \left|e_p^{\Upsilon^A} \otimes \prod_{i=0}^{N'-1}\varphi_i^{A1}\right\rangle \quad (16)$$

Eqn. 16 describes essentially a reversible purification of the AS by repeated interactions with auxiliary light pulses [10, 11]. In other word, if, while coaxing a quantum system to an eigenstate, Alice can record all the auxiliary states, and then q-teleport or transmit these states to Bob, he can accordingly (to Eqn. 15) reconstruct the original sate $|AS\rangle = |\Upsilon_0^A\rangle$ by performing a "reversed purification" on his own quantum system, BS, that has been initially prepared with an appropriate eigenstate state $|e_p^{\Upsilon^B}\rangle$. I will discuss the exact nature of such a purification process in relation to STT in the final section of this paper.

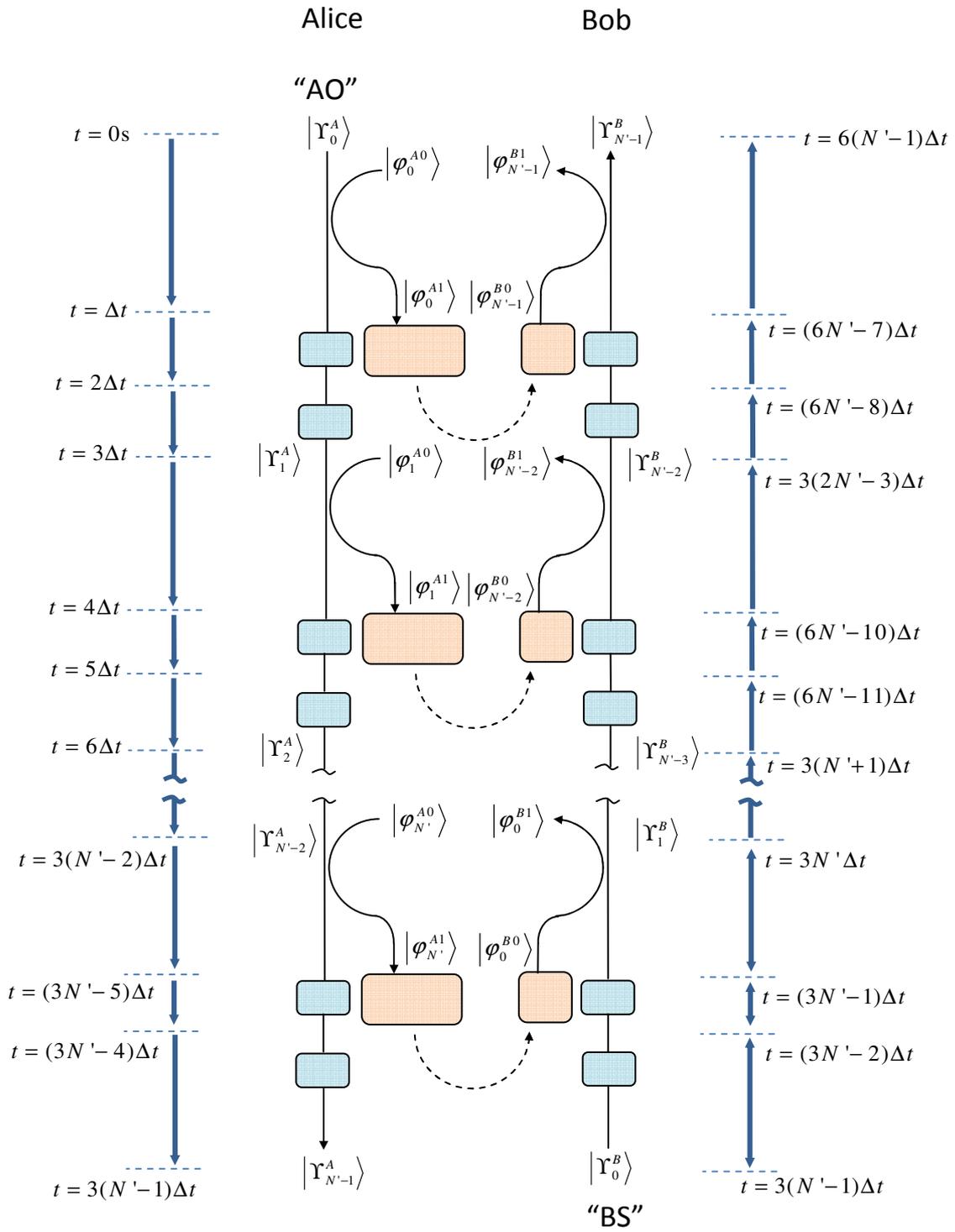

FIG. 2. General scheme for concatenated teleportation

## III. DIFFERENT TYPES OF CONCANATED TELEPORTATION

I shall now intend to classify a concatenated teleportation process. As will be clear shortly, the type of a concatenated teleportation process determines pivotally the minimum size of the quantum memory required.

Let's define a permutation operator $\hat{P}_{i,j}$ that swaps any two states $|\varphi_i\rangle$ and $|\varphi_j\rangle$ in $\left|\Upsilon \otimes \prod_{i=0}^{N'-1} \varphi_i \right\rangle$, e.g.

$$\hat{P}_{1,2}|\Upsilon \otimes \varphi_0 \otimes \varphi_1 \otimes \varphi_2 \otimes ... \varphi_{N'-1}\rangle = |\Upsilon \otimes \varphi_0 \otimes \varphi_2 \otimes \varphi_1 \otimes ... \varphi_{N'-1}\rangle \tag{17}$$

It can be seen that there exists two types of concatenated teleportation process:

**I. Type-1 concatenated teleportation,** where

$$\left(\hat{U}(\Delta t)^{-1}{}_{N'} \hat{R}^{-1}\hat{P}_{i,j}\right)^{N'} = \left(\hat{U}(\Delta t)^{-1}{}_{N'} \hat{R}^{-1}\right)^{N'}, i \neq j, \quad \forall i, j = 0, ..., N'-1 \tag{18}$$

**II. Type-2 concatenated teleportation,** where

$$\left(\hat{U}(\Delta t)^{-1}{}_{N'} \hat{R}^{-1}\hat{P}_{i,j}\right)^{N'} \neq \left(\hat{U}(\Delta t)^{-1}{}_{N'} \hat{R}^{-1}\right)^{N'}, i \neq j, \quad \forall i, j = 0, ..., N'-1 \tag{19}$$

Type-1 concatenated teleportation is possible only if the Hamiltonian $\hat{H}_\Upsilon \otimes \hat{H}_\varphi$ is decomposable as,

$$\hat{H}_\Upsilon \otimes \hat{H}_\varphi = \sum_{i=0}^{N'} \hat{H}_\Upsilon^i \otimes \hat{H}_\varphi^i \tag{20}$$

in which, $\hat{H}_\Upsilon^i \otimes \hat{H}_\varphi^i$ are non-trivial Hamiltonians, and obey the following commutation,

$$\left[\hat{H}_\Upsilon^i \otimes \hat{H}_\varphi^i, \hat{H}_\Upsilon^j \otimes \hat{H}_\varphi^j\right] = 0, \quad \text{iff } i = j$$

This is the simplest form of concatenated teleportation process that does not require large quantum memory. In fact, in a Type-1 concatenated teleportation, each of the partial teleportation steps (see Fig. 2) can be considered as an independent event, and thus the sequence of which is not entirely important. As such, Type-1 concatenated teleportation requires Alice and Bob each to possess a quantum memory that is only large enough to store a single light-pulse state $|\varphi^A\rangle$ (or $|\varphi^B\rangle$); the memory can be re-cycled after each partial teleportation event. Qiang Zhang et al.'s, and M. Al-Amri, et al.'s works in the teleportation of photonic multi-dimensional qubits composite systems [12, 13] are two examples of Type-1 concatenated teleportation.

However, unlike its photonic counterpart, the total Hamiltonian for the combined state $|\Upsilon \otimes \prod \varphi\rangle$ is generally not decomposable because the constituent fermions interact not only with the auxiliary photons, but also among themselves. In other word, the expression,

$$\hat{H}_\Upsilon \otimes \hat{H}_\varphi = \hat{H}_\Upsilon \otimes \hat{I} + \hat{I} \otimes \hat{H}_\varphi + \hat{H}_{\Upsilon \otimes \varphi}^{int} \qquad (21)$$

cannot generally be decomposed into $N'$ commutable Hamiltonians. Hence, STT must be a Type-2 concatenated teleportation, and a huge quantum memory would therefore be required so as to store the sequence of the states of the light pulses, $|\varphi_0^{A1}\rangle, |\varphi_1^{A1}\rangle, |\varphi_2^{A1}\rangle, ..., |\varphi_{N'-1}^{A1}\rangle$ (see Fig. 2).

## IV. STAR-TREK TELEPORTATION

I shall now apply Eqn. 15 to the simplest case of STT as illustrated in Fig. 3. Let's assumed that Bob and Alice are placed inside a 3D ellipsoidal reflector, their respective frame of reference ($\mathbf{L}$ and $\mathbf{L'}$, see Fig. 3) are mirror images of each other, and separated by some distance $d$. The origins of both frames are situated at the two foci of the reflector. The AS to be teleported is magnetically trapped at the focus $\mathbf{P}$ at Alice's end. A shutter $\mathbf{S}$ situated on the z-axis in $\mathbf{L}$ will open only momentarily at $t = 0, 3\Delta t, 6\Delta t, ..., 3(N'-1)\Delta t$ to allow just one auxiliary light pulse at a time to enter and interact with the AS at $\mathbf{P}$ in the cavity, but close otherwise. Thus, any light pulse originated from the AS is always reflected, and directed (green rays in Fig. 3) to a broadband high-fidelity parallel multimode first-in-last-out quantum memory (FILO-QMM), which divides the cavity of the reflector. It shall be assumed that the thickness of the FILO-QMM is extremely thin compared to the major diameter of the reflector, so that the paraxial approximation is applicable here [14]. The purpose of the FILO-QMM is to allow Bob to receive Alice's auxiliary pulses in reversed sequence. Although in this particular scenario, the same memory is shared between Alice and Bob, it suffices to demonstrate the STT concept, as can be seen later. For simplicity, I shall assume that the beam width of the incoming auxiliary pulses be comparable to the size of the AS. Bob possesses a setup similar to Alice's, except that it is a mirror image of the latter. Bob also traps his BS magnetically (at the focus $\mathbf{P'}$). The shutter $\mathbf{S'}$ basically operates in a reverse manner as $\mathbf{S}$, allowing light pulses originated from the BS to exit the cavity.

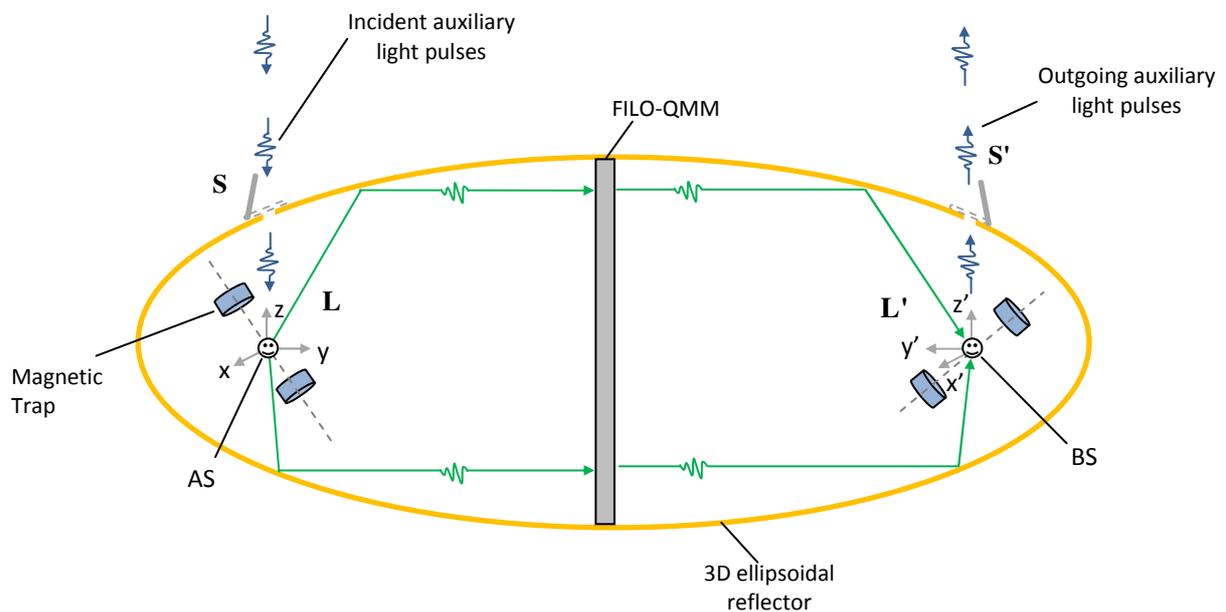

FIG. 3. Star-Trek Teleportation

Supposed Alice wishes to teleport both the q-information as well as the structural information about the AS situated at the focus $\mathbf{P}$, to Bob. I shall assume that AS consists solely of a gas of molecules. Note that the temperature of the AS needs not be very cold or near-condensation. To commence the teleportation process, Alice must first carry out a reversible purification process as pointed out in section II (see Eqn. 16). She does so by allowing a train of *N'* light pulses to impinge on the AS, via the shutter **S**, at a repetition rate of $1/(3\Delta t)$ (see Fig. 2). It will be assumed that she has carefully chosen the q-states of the light pulses, such that the sequence $|\varphi_0^{A0}\rangle, |\varphi_1^{A0}\rangle, |\varphi_2^{A0}\rangle, ...., |\varphi_{N'-1}^{A0}\rangle$ can effect a purification of the state $|AS\rangle$. The question here is into what eigenstate should $|AS\rangle$ be purified. Alice must consider a number of criteria for her choice. Firstly, in order to simplify the teleportation process, this particular state must be pre-agreed with Bob *a priori*, so that a classical measurement on the final purified state of the AS can be omitted. Secondly, it must be a state that Bob can easily prepare at his side. Thirdly, it must be a state that can be achieved by optical means (i.e. via the sequence of photon pulses, $|\varphi_0^{A0}\rangle, |\varphi_1^{A0}\rangle, |\varphi_2^{A0}\rangle, ...., |\varphi_{N'-1}^{A0}\rangle$). Lastly, it must be independent of the initial state of the AS. Here, I propose that this particular state should ideally be an atomic-Bose-Einstein-Condensate (BEC) state.

Hence, assuming that a density matrix $\hat{\rho}_o$ corresponding to the initial state $\left| AS \otimes \prod_i^{N'-1} \varphi_i^{A0} \right\rangle = \left| \Upsilon_0^A \otimes \prod_i^{N'-1} \varphi_i^{A0} \right\rangle$ be,

$$\hat{\rho}_o = \sum_{i,j=1}^{N_\Upsilon, N_\varphi} \sum_{m,n}^{N_\Upsilon, N_\varphi} \alpha_i^* \alpha_m \beta_j^* \beta_n \left| e_m^{\Upsilon^A} \right\rangle \left\langle e_i^{\Upsilon^A} \right| \otimes \left| e_n^{\varphi^A} \right\rangle \left\langle e_j^{\varphi^A} \right| \tag{22}$$

From Eqn. 22, the density matrix for $|AS\rangle$ is thus,

$$\hat{\rho}_{\Upsilon^A} = \text{Tr}_{|\Phi^A\rangle} \{\hat{\rho}_o\}$$

If the repeated interrogations on $|AS\rangle$ lead to a reversible cooling process, i.e.

$$\lim_{N' \to \infty} \text{Tr}\left\{\hat{H}_{\Upsilon^A} \text{Tr}_{|\Phi^A\rangle}\left(\hat{O}_A^{N'} \hat{\rho}_o \hat{O}_A^{-N'}\right)\right\} \to E_{\text{cond.}} \tag{23}$$

, where $E_{cond.}$ is the energy of the atomic-BEC state. Let the BEC state be $\left|e_p^{\Upsilon^A}\right\rangle$, one can conclude that, for a sufficiently large $N'$,

$$\lim_{N'\to\infty} \hat{O}_A \left|\Upsilon_0^A \otimes \prod_{i=0}^{N'-1} \varphi_i^{A0}\right\rangle \to \left|e_p^{\Upsilon^A} \otimes \prod_{i=0}^{N'-1} \varphi_i^{A1}\right\rangle \tag{24}$$

In other word, the condition prescribed by Eqn. 16 for a successful teleportation can thus be practically realized if AS can be optically cooled by a train of light-pulses. In order to render the cooling (or purification) process reversible and controllable, Alice can rely on Λ-type two-photon transitions when evolving the AS to the targeted eigenstate. To obtain the final atomic-BEC state, Alice can divide the purification process into two different parts. In the first, the molecular AS is decomposed into its constituent atoms via stimulated photo-dissociations (*dematerialization*) with a series of light pulses (entered via the shutter **S**) of appropriate photon-energies [15-17]. While doing so, she registers the q-states of all the light pulses (both the scattered and the stimulatedly-emitted photons) returned from the AS in the FILO-QMM. In the second part, she initiates another train of light pulses to cool the resultant atomic-AS to a condensate state. Again, while doing so, she registers the states of all the photons (both the scattered and the stimulatedly-emitted photons) returned from the atomic-AS in the FILO-QMM. The cooling process must be lossless and made reversible, say, by employing a combination of Λ-type transitions with circularly-polarized light, linearly-polarized light and a Sisyphus cooling scheme [18-21]. Alternatively, to assist the cooling process, Alice can first embed the molecules of interest in a BEC gas before loading the combined system into the magnetic trap at **P**. Preferably, the BEC gas should be inert to the spectral energies ($\hbar\omega$) used for the photo-dissociation (PD) process. The constituent atoms resulted from the PD can then be cooled sympathetically via cooling of the BEC host using light pulses with photon-energies differed from those used in the PD. In this case, the BEC host and the AS may be of different atomic species. Upon completion of the purification process, Alice releases the light pulse train stored in the FILO-QMM, in reversed sequence, to Bob.

Before proceeding to discuss how Bob can reconstruct the original molecular AS. The concept of equivalency between the Alice and Bob's eigenstates (e.g. $|e_i^{\Upsilon^A}\rangle$ and $|e_i^{\Upsilon^B}\rangle$) must be re-stated in the context of STT. According to Fig. 3, $|e_i^{\Upsilon^A}\rangle$ and $|e_i^{\Upsilon^B}\rangle$ (resp. $|e_i^{\varphi^A}\rangle$ and $|e_i^{\varphi^B}\rangle$) shall be considered equivalent if $\langle \bar{x}|e_i^{\Upsilon^A}\rangle = e_i^{\Upsilon^A}(\bar{x}) = \langle \bar{x}'|e_i^{\Upsilon^B}\rangle = e_i^{\Upsilon^B}(\bar{x}')$ (resp. $\langle \bar{x}|e_i^{\varphi^A}\rangle = e_i^{\varphi^A}(\bar{x}) = \langle \bar{x}'|e_i^{\varphi^B}\rangle = e_i^{\varphi^B}(\bar{x}')$), where $\bar{x}$ and $\bar{x}'$ are vectors defined with respect to $\mathbf{L}$ and $\mathbf{L'}$, respectively.

To reconstruct the original state $|\Upsilon_0^A\rangle$ of the AS, Bob must, in principle, carry out the inverse unitary operation described in Eqn. 15. This, however, is realistically impossible as it will require Bob to reverse the arrow of time. Nonetheless, Bob can still "emulate" the time-reversal process with the (forward) time-evolution operators, $\hat{U}_{\hat{H}_\Upsilon \otimes \hat{H}_\varphi}$ and $\hat{U}_{\hat{H}_\Upsilon}$, by performing on the state $|BS\rangle$ and the recalled state $|\varphi_i^{B0}\rangle$s from his quantum memory, a specific symmetric transformation $\hat{t}$ (see Fig. 2).

To see how this can be achieved, I note that the AS is a time-reversal invariant system, and hence the Eugene Wigner's time-reversal theorem is applicable here [22],

$$e^{i(\hat{T}(\hat{H}_\Upsilon \otimes \hat{H}_\varphi)\hat{T}^{-1})(\Delta t')}\left(\hat{T}|\Upsilon(-\Delta t') \otimes \varphi(-\Delta t')\rangle\right) = \hat{T}e^{-i(\hat{T}(\hat{H}_\Upsilon \otimes \hat{H}_\varphi)\hat{T}^{-1})(\Delta t')}|\Upsilon(-\Delta t') \otimes \varphi(-\Delta t')\rangle$$
$$= \hat{T}\hat{U}^{-1}(\Delta t)|\Upsilon(\Delta t) \otimes \varphi(\Delta t)\rangle \quad (25)$$
$$= \hat{T}|\Upsilon(t=0) \otimes \varphi(t=0)\rangle$$

where $\Delta t' = -\Delta t$ and $|\Upsilon(-\Delta t') \otimes \Phi(-\Delta t')\rangle = |\Upsilon(\Delta t) \otimes \Phi(\Delta t)\rangle = \hat{U}(\Delta t)|\Upsilon(t=0) \otimes \Phi(t=0)\rangle$,

and $\hat{T} = \hat{T}_\Upsilon \otimes \hat{T}_\varphi$ is the combined Wigner's time-reversal operator acting on $|\Upsilon\rangle$ and $|\varphi\rangle$.

Now, I define a reversal operator, $\hat{t}$ ($= \hat{t}_\Upsilon \otimes \hat{t}_\varphi$), as

$$\hat{t} = \hat{T}\hat{K} = \hat{K}\hat{T} = \hat{T}_\Upsilon \hat{K}_\Upsilon \otimes \hat{T}_\varphi \hat{K}_\varphi = \hat{K}_\Upsilon \hat{T}_\Upsilon \otimes \hat{K}_\varphi \hat{T}_\varphi$$

where $\hat{K}$ is the conjugation. It is clear that $\hat{t}$ is a symmetry transformation; $\hat{t}\hat{p}\hat{t} = -\hat{p}$, $\hat{t}\hat{l}\hat{t} = -\hat{l}$, etc, and $\hat{t}i\hat{t} = i$.

From Eqn. 25, we can see that,

$$e^{i(\hat{T}(\hat{H}_\Upsilon \otimes \hat{H}_\varphi)\hat{T}^{-1})(\Delta t')} \left(\hat{T}|\Upsilon(-\Delta t') \otimes \varphi(-\Delta t')\rangle\right) = \hat{T}|\Upsilon(t=0) \otimes \varphi(t=0)\rangle$$
$$\hat{t}e^{-i(\hat{H}_\Upsilon \otimes \hat{H}_\varphi)(\Delta t')}|\Upsilon(-\Delta t') \otimes \varphi(-\Delta t')\rangle = \hat{t}|\Upsilon(t=0) \otimes \varphi(t=0)\rangle$$
$$e^{-i(\hat{H}_\Upsilon \otimes \hat{H}_\varphi)(\Delta t)}\hat{t}|\Upsilon(\Delta t) \otimes \varphi(\Delta t)\rangle = \hat{K}\hat{t}|\Upsilon(t=0) \otimes \varphi(t=0)\rangle \quad (26)$$
$$\hat{U}(\Delta t)\left(\hat{t}|\Upsilon(\Delta t) \otimes \varphi(\Delta t)\rangle\right) = \hat{K}\left(\hat{t}|\Upsilon(t=0) \otimes \varphi(t=0)\rangle\right)$$

Hence, I can replace $\hat{O}_B$ in Eqn. 15 with $\hat{O}'_B \hat{t}$, and re-write the equation as,

$$\left|\Upsilon^B_{N'} \otimes \prod_{i=0}^{N'-1}\varphi^{B1}_i\right\rangle = \left(\hat{O}'_B \hat{t}\hat{M}_{B,M\to\varphi}\right)\hat{P}_p \hat{T}_{A\to B}\left(\hat{M}_{A,\varphi\to M}\hat{O}_A\right)\left|\Upsilon^A_0 \otimes \prod_{i=0}^{N'-1}\varphi^{A0}_i\right\rangle \quad (27)$$

in which, $\hat{O}'_B = {}_{N'}\hat{R}\hat{U}_{\hat{H}_{\Upsilon B} \otimes \hat{H}_{\varphi B}}(\Delta t)\left(\hat{U}_{\hat{H}_{\Upsilon B}}(2\Delta t) \otimes \hat{I}\right)\left(\hat{t}_\varphi {}_{N'}\hat{R}\hat{U}_{\hat{H}_{\Upsilon B} \otimes \hat{H}_{\varphi B}}(\Delta t)\left(\hat{U}_{\hat{H}_{\Upsilon B}}(2\Delta t) \otimes \hat{I}\right)\right)^{N'-1}$.

Note that the expression for $\hat{O}'$ does not contain any inverse unitary term $\hat{U}^{-1}$. From Eqn. 27, it thus follows that if Alice can perform a unitary described in Eqn. 24 on AS to obtain a BEC state $|e^{\Upsilon^A}_p\rangle$, Bob should be able to reconstruct the original $|\Upsilon^A_0\rangle$ state by first preparing his own N-body fermionic system (of same atomic species as the AS) with a BEC state $|BS\rangle = |e^{\Upsilon^B}_p\rangle$, and then reversing the cooling (and the photo-dissociation) process using the sequence of light-pulse states ($|\varphi^{A1}_0\rangle, |\varphi^{A1}_1\rangle, |\varphi^{A1}_2\rangle, ..., |\varphi^{A1}_{N'-1}\rangle$) he has received from Alice. To elicit a "time reversal", Bob simply performs on each of these states a reversal transformation $\hat{t}_\varphi$. Fortunately, the use of a 3D ellipsoidal reflector automatically effects such an operation at Bob's side. Additionally, he should act upon his own condensate ($\left|e_p^{\Upsilon^B}\right\rangle$) a reversal transformation $\hat{t}_\Upsilon$. Since a trapped condensate generally possesses no momentum, $\hat{t}_\Upsilon$ can be approximated as an identity for a BEC state, i.e. $\left|e_p^{\Upsilon^B}\right\rangle \approx \hat{t}_\Upsilon \left|e_p^{\Upsilon^B}\right\rangle$

We can therefore see that Eqn. 27 should lead to,

$$\left|\Upsilon_{N'}^B \otimes \prod_{i=0}^{N'-1} \varphi_i^{B1}\right\rangle = \left(\hat{Q}_\Upsilon \otimes \hat{Q}_\varphi\right)\hat{K}\left|\Upsilon_0^A \otimes \prod_{i=0}^{N'-1} \varphi_i^{A0}\right\rangle$$

or rather,

$$\left\langle \bar{x}_\Upsilon' \middle| \Upsilon_{N'}^B \right\rangle = \left\langle \bar{x}_\Upsilon \middle| \hat{K}_\Upsilon \middle| \Upsilon_0^A \right\rangle = \Upsilon_{N'}^B(\bar{x}_\Upsilon') = \Upsilon_0^A(\bar{x}_\Upsilon)^*$$

The final equation implies that Bob can re-create (*rematerialisation*) the physical structure of the AS ($\Upsilon_0^A(\bar{x}_\Upsilon)$) without the need to reverse the time-arrow, i.e. without having had to perform the inverse operations $\hat{U}^{-1}_{\hat{H}_{\Upsilon^B} \otimes \hat{H}_{\varphi^B}}(\Delta t)$ and $\hat{U}^{-1}_{\hat{H}_{\Upsilon^B}}(2\Delta t) \otimes \hat{I}$. Note that however the reconstructed object will be an mirror image of the original, since **L'** is a mirror image of **L**. If Bob is very stringent about the symmetricity of the final object, a double STT process via an intermediate party, say Victor, would be necessary, so that the sample can be inverted back to its original symmetry at Bob's side.

## V. CONCLUSION

I have shown that a time-reversal invariant system can be teleported in a "Star-Trek" manner. Owing to the complexity of the AS involved, STT is a type-2 concatenated teleportation process and a broadband large-capacity FILO-QMM is required. Such a memory can likely be constructed based on the technique recently reported by Mahdi Hosseini, et al. [23]. My subsequent papers will look into various issues affecting the fidelity of the STT. These include quantum chaos and de-coherence due to vacuum fluctuations.